# Principle of Least Action and Theory of Cyclic Evolution

Atanu Bikash Chatterjee

Department of Mechanical Engineering, Bhilai Institute of Technology, Bhilai house
Durg, 490006 Chhattisgarh, India
e -mail: abc3.14160@yahoo.com

**Abstract:** A natural process is defined as an act, by which a system organizes itself with time. Any natural process drives a system to a state of greater *organization*. *Organization* is a progressive change, while evolution is expressed in the effects of accumulating marks acquired from contingent encounters. Co-existence of the system in states of maximum *organization* as well as maximum *action* forms the core idea of the paper. Major influences have been drawn from the Principle of Least Action. This allows us to see how this most basic law of physics determines the development of the system towards states with less *action* *i.e.* organized states. Based on this, it has been proposed, that the development of a system towards states of greater *organization* is *cyclic* in nature and thus evolution is a *cyclic* process.

**Introduction:** *Organization*[1,2] is a progressive change and can be modelled as a part of nature. Nature comprises of open systems. An open system is a continually evolving dynamical system. All natural processes[3,4] occurring in the universe are rooted in physics and have physical explanation. All of the structures in the universe exist, because they are in their state of least *action*[5] or tend towards it. In any system, simple or complex, the system spontaneously calculates which path will use least effort for that process[6,7]. A system comprises of elements and constraints, both internal as well as external. The internal constraints could be the configurations of the system or the state of elements themselves, whereas, the external constraints are those that define the geometry of the system. The elements apply work on the constraints to modify the *organization* and minimize the *action*, which takes finite amount of time, making the reorganization a process[6,7] Reorganization is a process of optimization. A system thrives to organize itself and in the course of development destroys its previous identity. The dynamical systems that are present in nature are generally very complex exhibiting various levels of *complexity* present within themselves. Order implies a state of lesser *action* hence, greater *organization*. A complex system with a structure and emergence is said to self-organizing[1,2]. The process of self-organization of the systems can be called a "Process of achieving a least action state by a system". It could last billions of years or indefinitely.

$$i \xrightarrow{X_t} f_t$$

Figure representing an evolving system undergoing a natural process ($X_t$) in time ($t$) from a state of lesser organization ($i$) to a state of greater organization ($f_t$).

Evolution of a system towards a greater state of *organization* is a coherent *action* of the *organization* of each system elements towards their optimal states of *organization*[8]. The extent of *organization* achieved by a system element depends upon its work potential or its *exergy*[7, 9] compared to its surrounding media. This energy gradient acts as a driving force enabling a dynamical system to organize itself with continuous evolution of time. In an open system, there is always an influx and out flux of energy between the system and surrounding media, causing the *exergy* of the system elements to vary continuously. So, the *action* of a single element will not be at minimum, but the sum of the *action* of all the elements in the system will be at minimum. The *action* of a single element is not maximal as well, because by definition this will destroy the system, so this intermediate state represents an optimum[6, 10].

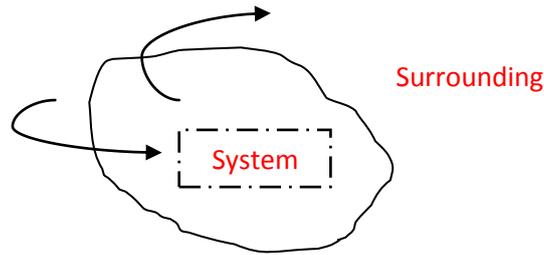

Figure representing an exchange of energy between a system and the surrounding

The extent of *organization* achieved by a system element depends upon its work potential or its *exergy* compared to its surrounding media. In an open system there is always an influx and out flux of energy between the system and surrounding media, causing the *exergy* of the system elements to vary continuously. This exchange of energy is accompanied with exchange in *entropy* between system and surrounding [11, 12, 13, 14]. The sum total of the *entropy* generated within the system and the *entropy* exchanged gives the *entropy* generated in a natural process.

$$(\partial S)_{gen} = (\partial S)_{int} + (\partial S)_{ex} \qquad (1)$$

Where, $(\partial S)_{gen}$ represents total *entropy* generated[8], $(\partial S)_{int}$ represents total internal *entropy* of the process and $(\partial S)_{ex}$ represents total exchange of *entropy* between system and surrounding media.

$$(\partial S)_{ex} = |(\partial S)_{in}| + |(\partial S)_{ex}| \qquad (2)$$

Where, $(\partial S)_{ex}$ is the sum total of the influx and out flux of *entropy*. A natural process is also accompanied by the increase in number of microstates of the system. Thermodynamically, *entropy* is simply our lack of knowledge of the actual state of the system[8]. Thus, with increase in time and hence, increase in *organization* the system elements tend to lose track of their evolutionary history. The lack of information with increasing *organization* thus, renders a system towards greater levels of *complexity*[8, 15].

$$(\partial S)_{int} > 0$$

Let a system be initially in a state '*i*' and make a transition into the $t^{th}$ final state '$f_t$' through a natural process $X_t$ causing an increase in the amount of *organization*, where '*t*' represents the time elapsed while undergoing the process and $t \in (0, \infty)$.

The final state of the system is unknown since the system under consideration is open to surrounding media[5] hence; it has been subscripted with '*t*'. Thus, the final fate of the system can assume any out of the infinite states, $f_t = f_0, f_1, f_2, f_3 \ldots\ldots\ldots f_\infty$

Where, $f_0$ is same as the initial state '*i*'.

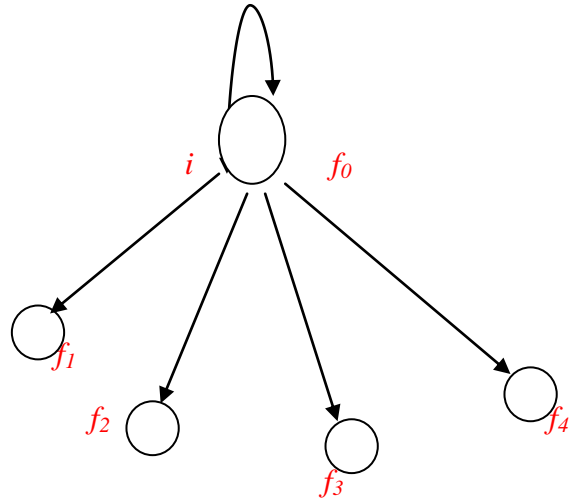

Figure representing transition diagram for a system

The figure represents the transition diagram for a system from $i^{th}$ state to $f^{th}$ state where, transition from *i* to $f_1$ is denoted by the natural process $X_1$ and so on. Each process can occur in an infinite number of ways, thus, leading the system towards infinite number of final states. But the Principle of Least Action[5, 6, 7] imposes constraint, by causing the system to undergo a specific process out of the available infinite processes. According to the Principle of Least Action, any natural process occurs in that way which consumes the least time. The transition of a system from one state to another is a coherent phenomenon of all its constituting elements. An orderly motion at the microscopic level of the system causes a visible motion at the macroscopic level[8]. The amount of *organization* present in the system at a later state is always greater than the amount of *organization* it possessed at an earlier state. Increasing the amount of *organization* is the driving force behind any natural process. However, the *rapidity* involved in a natural process depends upon the time each of the system elements take to organize themselves. The time taken by each element to organize itself varies continuously because, the system is always communicating with the surrounding media by exchanging energy and *entropy*[8, 11]. So each system element possesses a set of strategy[16, 17, 18]. *Strategies for a system element are its trajectories in phase space obtained from solving its integral equation of motion*. The set of strategies for an element is called pure strategy if the system is free from any constraints. Presence of constraints causes the system elements to optimize their strategies in order to follow the least path and organize themselves[9]. Optimization thus, prevents a system element to use its pure strategy. Reorganization of a system or its constituting elements is thus a process of optimization. In presence of constraints, the set of strategies thus, employed by the system elements are their mixed strategies.

The Principle for Least Action states that the actual motion of a conservative dynamical system between two points, occurs in such a manner, that the *action* has a minimum value in respect to all other paths between the points, which correspond to the same energy[5, 6].

The classical definition of the Principle of Least Action[5] is:

$$I = \int_{t_1}^{t_2} L\,dt \quad (3)$$

$$L = T - V$$

Where, *I* is the *action* of the system, *L* is the Lagrangian of the system and *T* and *V* are the kinetic and the potential energy of the system respectively. The variation of the path is zero for any natural process occurring between two points of time $t_1$ and $t_2$, or the nature acts in the simplest way hence, in the shortest possible time.

For the motion of the system between time $t_1$ and $t_2$, the Lagrangian, *L*, has a stationary value for the correct path of motion.

$$\delta I = \delta\left(\int_{t_1}^{t_2} L\,dt\right) = 0 \quad (4)$$

Eqn. (4) can be summarized as the Hamilton's Principle[5].

For a system consisting of *N*-elements,

$$L = \sum_{j=1}^{N} L_j \quad (5)$$

$$\text{where, } L_j = T_j - V_j$$

Where, $L_j$, $T_j$ and $V_j$ represent the Lagrangian, kinetic and potential energies of the $j^{th}$ system elements.

So,

$$I = \sum_{j=1}^{N} I_j = \int_{t_1}^{t_2} \sum_{j=1}^{N} L_j\,dt = \int_{t_1}^{t_2} \sum_{j=1}^{N}(T_j - V_j)\,dt \quad (6)$$

Where, $I_j$ is *action* of the $j^{th}$ system element.

Let the set of pure strategies [17, 18] for the $j^{th}$ element, corresponding to the $t^{th}$ final state be given by, $\pi_{f_t}^{j}$

$$\text{where, } \pi_{f_t}^{j} = \left(\pi_{f_0}^{j}, \pi_{f_1}^{j}, \pi_{f_2}^{j}, \ldots\ldots\ldots\ldots\right), \forall t \in (0, \infty) \quad (7)$$

Let $p_j$ be a continuous function that maps the set of all n-tuples of pure strategies for each element into real numbers. These sets of real numbers form the set of mixed strategies for each element. Let the set of pure strategies for the $j^{th}$ element, corresponding to the $t^{th}$ final state be given by, $\mu_{f_t}^{j}$

$$\text{so, } p_j\left(\pi_{f_t}^{j}\right) = \mu_{f_t}^{j}, \forall t \in (0, \infty) \quad (8)$$

Eqn. (8) is subjected to the constraints;

$$\mu_{f_t}^j \geq 0 \; \forall j \in (1, N) \; and \; \sum_{t=0}^{\infty} \mu_{f_t}^j = 1 \; \forall t \in (0, \infty) \tag{9}$$

The constraints of the process validate the occurrence of the natural process, since the sum of the mixed strategies for a system element and the probability of occurrence of a process is equal to unity. From the above conditions presented in eqn. (8) and eqn. (9) it can be clearly observed that, $p_j$ is simply a probability density function[19] that operates on the random variable $\pi_{f_t}^j$. The mixed strategy for the system at the macroscopic level is denoted by $\mu_{f_t}$.

$$\mu_{f_t} = \sum_{j=1}^{N} \mu_{f_t}^j, \forall j \in (1, N) \tag{10}$$

For the occurrence of the phenomena at the macroscopic level, *coherence* in microscopic level must be existent[8].

$$\sum_{t=0}^{\infty} \mu_{f_t} = \sum_{t=0}^{\infty} \sum_{j=1}^{N} \mu_{f_t}^j = 1, \forall t \in (0, \infty) \, and \, j \in (1, N) \tag{11}$$

A system's mixed strategy is thus, a probability distribution. A system's pay-off represents the amount of *organization*, $(Org)$ it posses[17, 18]. Let $\rho_j$ be the pay-off function that maps the set of mixed strategies for each set of the system elements and in turn generates the pay-off for each system element, which is denoted by;

$$\rho_j\left(\mu_{f_t}^j\right) = (Org)_{f_t}^j = \frac{1}{I^j} = \frac{1}{\int_0^t L^j \, dt}$$

The strategy chosen by the system element is that, which tends to maximize the amount of *organization* present within the system, at a later state. The set of optimal strategy for the $j^{th}$ system element that maximizes the amount of *organization* of the system is thus, a process of optimization[6, 7, 10]. The system elements evolve with time and achieve their set of optimal strategies that maximizes *organization* of the system as a whole. This set of strategies for a $j^{th}$ system element represents its *Nash equilibrium* strategy[17, 18] profiles, denoted by $\hat{\mu}_{f_t}^j$.

$$so, \sum_{j=1}^{N} \rho_j\left(\hat{\mu}_{f_t}^j\right) = max(Org)_{f_t} \; at \; a \; particular \; time \tag{12}$$

An organized system tends to have the least value of *action*. Conversely, lesser is the value of *action* more is the amount of *organization* present in a system[10]. Thus, amount of *organization (Org)* is inversely related to the *action* of the system *(I)*.

$$(Org) \times (I) = constant \tag{13}$$

Differentiating with respect to time[7],

$$\frac{\partial Org}{\partial t} = -\left(\frac{Org}{I}\right) \times \frac{\partial I}{\partial t} \tag{14}$$

The equation implies that, the rate of increase of *organization* in a system is equal to the rate of decrease in *action* of the system multiplied by the ratio of amount of *organization* to the amount of *action* possessed by the system at an earlier state.

So, the above equation can be rewritten as;

$$\left(\frac{\partial Org}{\partial t}\right)_{f_t} = -\left(\frac{Org}{I}\right)_i \times \left(\frac{\partial I}{\partial t}\right)_{f_t} \qquad (15)$$

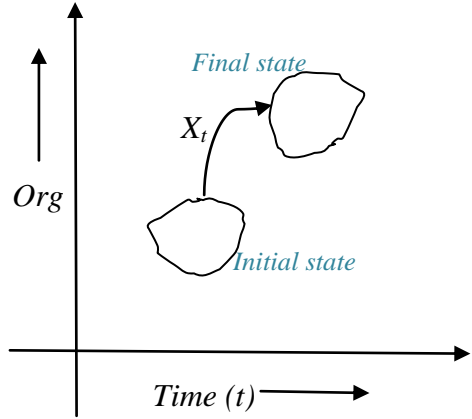

Figure representing a system undergoing a natural process from lesser organized initial state to greater organized final state

The rate of increase of *organization* is the *directionality* of a natural process. For any natural process the ratio of *organization* to *action* at an earlier state must always be greater than unity. This is because of the fact that a system would have ceased to exist at the earlier state if the ratio became less than unity.

From eqn. (15) we have,

$$\left(\frac{\partial Org}{\partial t}\right)_{f_t} = -\left(\frac{Org}{I}\right)_i \times \left(\frac{\partial I}{\partial t}\right)_{f_t}$$

An equation governing all natural processes[3, 4, 7, 20] can be presented as;

$$X_t = \left(\frac{\partial Org}{\partial t}\right)_{f_t} = -\left(\frac{Org}{I}\right)_i \times \left(\frac{\partial I}{\partial t}\right)_{f_t} \qquad (16)$$

$$X_t = -\alpha \left(\frac{\partial I}{\partial t}\right)_{f_t} \qquad (17)$$

The above result when modified for a system undergoing a natural process becomes,

$$X_t = -\alpha \left(\frac{\partial I}{\partial t}\right)_{f_t} = -\alpha \frac{\partial}{\partial t}\left(\int_0^t \sum_{j=1}^N L^j \, dt\right) \qquad (18)$$

$$\frac{\partial L^j}{\partial t} = -\frac{dh^j}{dt}$$

Here, $h^j$ represents energy function[5] of the $j^{th}$ system element and $h^j = T^j + V^j$.

So, eqn. (18) gets reduced to,

$$X_t = \alpha \int_0^t \left(\sum_{j=1}^N \frac{dh^j}{dt}\right) dt \qquad (19)$$

The physical significance of this equation is that systems undergoing natural process organize themselves with time and proceed with energy dispersal of their constituting elements [20, 21].

Let $r$ be defined as the rate of a natural process,

$$r = \frac{\partial X_t}{\partial t} = \alpha \frac{\partial}{\partial t}\left(\int_0^t \left(\sum_{j=1}^N \frac{dh^j}{dt}\right) dt\right)$$

*Rapidity (r)* associated with a natural process is a very important tool. *Rapidity* is also the rate of increase of *complexity* of a system. It can be used as a tool for comparing between the rates of evolution of two identical systems. Also, *rapidity* is a decreasing function with respect to time for any natural process. Every system element tends towards a state of greater *organization* by reducing its *exergy* compared to the surrounding media by dispersal of its free energy [20, 21]. Thus, the *rapidity* of a system decreases as the system evolves with time and proceeds towards the state of maximum *organization*.

*Rapidity* plays a very important role in the existence and evolution of the system as a whole. If individual system elements are assigned a unique magnitude of *rapidity*, denoted as $r^j$ where,

$r^j$ is the *rapidity* associated with the $j^{th}$ system element, then,

$$r \propto \min_{all\ j} r^j$$

The above proposition is a significant and very important result. The rate or *rapidity* of a system evolving through a natural process is directly proportional to its least rapid constituting element, *i.e.,* the system element that is least rapid in achieving a state of greater organization governs the overall *rapidity* of the system.

*If the action possessed by the system becomes zero[7]*:

$$I = \int_{t_1}^{t_2} L dt = 0$$

From eqn. (13) it can be easily seen that, when *action* of a system becomes zero the amount of *organization* becomes infinite. A system on achieving the state of maximum *organization* comes in equilibrium with the surrounding. The amount of *action* present within a system is a property of the system itself, it is determined by the state of its constituting elements and system constraints. When *action* assumes a null value, the system shrinks to a singular point. Approaching a more organized state is the natural tendency of a system. Independent of the instantaneous configuration, the system continuously reconfigures itself with time. But, when the *action* becomes zero, as in this case, either the system collapses by shrinking to an infinitesimal point or it attains a state of maximum *organization*. Attaining a state of maximum *organization* implies the disappearance of the energy gradient between the system and its surrounding. Such a system is said to have reached a *dead state*[9] where all natural processes have ceased to exist. On reaching the *dead state,* the system no longer evolves with time but becomes a static structure. The system would continue to remain at that state for an infinite period of time. A least *action* state is also a state of least amount of free energy[3, 20]. A system's configuration determines the amount of free energy it possesses. A system with zero *action* then must have no free energy and hence, no configuration. This implies that after achieving the *dead state* the system begins to shrink to a point[7], or more precisely both the processes occur almost simultaneously.

From eqn. 1 we have

$$(\partial S)_{gen} = (\partial S)_{int} + (\partial S)_{ex}$$

And from eqn. 2 we know

$$(\partial S)_{ex} = |(\partial S)_{in}| + |(\partial S)_{ex}|$$

Eqn.1 modifies into,

$$(\partial S)_{gen} = (\partial S)_{int} + (\partial S)_{in} - (\partial S)_{out} \qquad (20)$$

(*Influx of entropy is considered positive and out flux is considered to be negative.*)

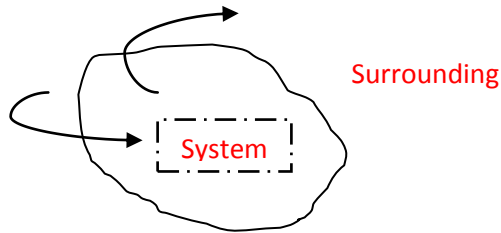

Figure denoting entropy exchange between system and surrounding

At equilibrium, $(\partial S)_{gen} = 0$, a necessary condition for evolution as systems in nature tend towards the state of maximum *organization*[11, 12, 13].

So, eqn.20 is rewritten as

$$(\partial S)_{gen} = (\partial S)_{int} + (\partial S)_{in} - (\partial S)_{out} = 0 \qquad (21)$$

At equilibrium the boundary separating the system and surrounding collapses and the total *entropy* generated within the system is flushed out to the surrounding. Influx of *entropy* thus loses its significance.

$$(\partial S)_{gen} = (\partial S)_{int} - (\partial S)_{out} = 0 \qquad (22)$$

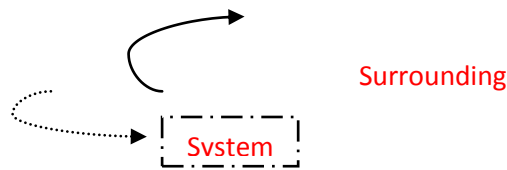

Figure representing the collapse of the system and system-surrounding boundary vanishes at state of equilibrium

Conversely, $\gamma$ be an instant between the time interval $(t_1, t_2)$. Since, *action* of a system is a continuous function of time the time interval can be represented as $((t_1, \gamma), (\gamma, t_2))$.

So, *action* can be represented as[7];

$$I = \int_{t_1}^{\gamma} L dt + \int_{\gamma}^{t_2} L dt$$

Eqn. (17) modifies into,

$$X_t = -\alpha \left(\frac{\partial I}{\partial t}\right)_{f_t} = -\alpha \left\{\frac{\partial}{\partial t}\left(\int_{t_1}^{\gamma} L dt + \int_{\gamma}^{t_2} L dt\right)\right\}_{f_t} \quad (23)$$

At equilibrium, natural processes cease to exist, so, $X_t = 0$

This implies that the quantity within the derivative assumes a stationary value.

So,

$$\int_{t_1}^{\gamma} L dt + \int_{\gamma}^{t_2} L dt = stationary\ value$$

But at equilibrium, *action* vanishes,

$$so, \int_{t_1}^{\gamma} L dt + \int_{\gamma}^{t_2} L dt = 0 \quad (24)$$

$$and \int_{t_1}^{\gamma} L dt = -\int_{\gamma}^{t_2} L dt$$

$$t_1 < \gamma < t_2$$

'$\gamma$' clearly denotes a point of inflexion for any natural process ranged over certain interval of time. In the time span, $(t_1, \gamma)$ *action* assumes a positive value and in the time span $(\gamma, t_2)$ it assumes a negative value or vice-versa. In the time interval $(t_1, \gamma)$ amount of increase in *organization* of a system is directly proportional to the amount of decrease in *action* whereas, in the time interval $(\gamma, t_2)$ the amount of increase in *organization* of a system is directly proportional to the amount of *increase* in *action* as well. The extent to which the system organizes in the first phase (time interval) gets destroyed in the second phase. $\gamma$, thus represents the instant where maximum *organization* and disorganization co-exist. At equilibrium state, *rapidity* has no significance.

At equilibrium,

$$X_t = \alpha \int_0^t \left(\sum_{j=1}^{N} \frac{dh^j}{dt}\right) dt = \tau \frac{\partial S_{int}}{\partial t} = 0$$

Here, $\tau$ is the constant of proportionality.

Hence, at equilibrium it is clearly seen that, all natural processes cease to exist and internal *entropy* becomes maximum[8, 9, 12]. The system becomes highly organized and exhibits maximum level of *complexity*. But, on the other hand the free energy of the system becomes maximal and the system instantly disintegrates.

**Conclusion:** Nature in its crude form is very difficult to understand but we must not get carried away by the simplicity of the laws that are thought to govern nature. As once pointed out by Feynman[7,22](1948), there is a pleasure in recognizing old things from a new point of view. *Action* being an extensive property first vanishes causing the system to get highly organized and then causes it to shrink into an infinitesimal point and then re-appears at its maximum magnitude causing the system to become highly unpredictable. Re-organization starts now in the disintegrated system causing it to develop its levels of *complexity*. This unpredictable system again tries to achieve the state of least *action* and the cycle continues forever. However, this time the course of its development may be entirely different.

*Every natural process passes through three stages of evolution: organization, disintegration and re-organization. Global Complexity occurs thus at the edge of chaos. Organization, disintegration and again reorganization is a cyclic process. At this state the system behaves chaotically and its future course of evolution is highly sensitive to initial conditions.* Natural games are not antagonistic in nature. For global existence, antagonistic games do exist in nature, *e.g.,* Predator-Prey models, etc. But at local microscopic level, system elements optimize their *action* in a coherent manner *co-operatively* for global maximization of *organization* at macroscopic level. *No open systems in nature can exist indefinitely. Desire of achieving greater organization ultimately drives these systems towards self-destruction.*